\documentclass[aps,prb,amsmath,amssymb,twocolumn]{revtex4}

\usepackage{bm}
\usepackage{graphicx}

\DeclareMathOperator{\Img}{\mathrm{Im}}
\DeclareMathOperator{\Rea}{\mathrm{Re}}
\DeclareMathOperator{\Sp}{\mathrm{Sp}}

\begin{document}
\title{Phase-coherent thermoelectricity in superconducting hybrids (Brief Review)}
\author{Mikhail S. Kalenkov}
\affiliation{I.E. Tamm Department of Theoretical Physics, P.N. Lebedev Physical Institute, 119991 Moscow, Russia}
\author{Andrei D. Zaikin}
\affiliation{Institute for Quantum Materials and Technologies, Karlsruhe Institute of Technology (KIT), 76021 Karlsruhe, Germany}
\affiliation{I.E. Tamm Department of Theoretical Physics, P.N. Lebedev Physical Institute, 119991 Moscow, Russia}

\begin{abstract}
We review some recent advances in studies of phase-coherent thermoelectric effects in superconducting hybrid structures such as, e.g., Andreev interferometers. We elucidate a number of mechanisms of electron-hole symmetry breaking in such systems causing dramatic enhancement of thermoelectric effects. We demonstrate that the flux-dependent thermopower exhibits periodic dependence on the applied magnetic flux $\Phi_x$ which in some limits may reduce to either odd or even functions of  $\Phi_x$ in accordance with experimental observations. We also show that dc Josephson current in Andreev interferometers can be controlled and enhanced by applying a temperature gradient which may also cause a nontrivial current-phase relation and a transition to a $\pi$-junction state.
\end{abstract}

\maketitle

\section{Introduction}
\label{sec:intro}
Applying an electric field $\bm{E}$ to a normal
conductor with Drude conductivity $\sigma_N$ one induces an electric current $\bm{j}=\sigma_N \bm{E}$ across this conductor. Likewise, such a current can be generated by applying a thermal gradient $\nabla T$, in which case one has $\bm{j}=\alpha_N \nabla T$. This simple equation illustrates the essence of the so-called thermoelectric effect in normal metals. 

Usually the latter effect turns out to be quite small since contributions to the thermoelectric coefficient $\alpha_N$ from electron-like and hole-like excitations are of the
opposite sign and almost cancel each other. As a result, $\alpha_N$ is proportional to a small ratio between temperature $T$ and the Fermi energy $\varepsilon_F$, i.e. one has $\alpha_N \sim (\sigma_N/e)(T/\varepsilon_F)$. 

This simple physical picture gets substantially modified as soon as a normal metal is brought into a superconducting state.
On one hand, the electric field no longer penetrates into a superconductor and, hence, the
Drude contribution to the current vanishes. On the other hand, a  supercurrent $\bm{j}_s$ can now be induced in the
system without applying any electric field.  This current  exactly compensates for any thermoelectric current, $\bm{j}_s =- \alpha_S \nabla T$,
thus making the net electric current vanish. For this reason the thermoelectric effect in uniform superconductors cannot be detected 
in any way.

The way out has been suggested by Ginzburg \cite{Ginzburg44,Ginzburg91} who has demonstrated that no such compensation takes place
in spatially non-uniform superconductors. This observation offers a possibility to experimentally investigate thermoelectric currents 
in such structures as, e.g., superconducting hybrids. Still, similarly to normal metals the magnitude of the thermoelectric effect in superconductors was expected \cite{Galperin73,Aronov81} to be very small, i.e. $\alpha_S \sim \alpha_N$ at $T \sim T_C$ and $\alpha_S \ll \alpha_N$ at $T \ll T_C$, where $T_C$  is the superconducting critical temperature.

Quite unexpectedly, already first experiments with bimetallic superconducting rings
\cite{Zavaritskii74,Falco76,Harlingen80} revealed the thermoelectric signal which magnitude 
and temperature dependence were in a strong disagreement
with theoretical predictions \cite{Galperin73,Aronov81}. In particular, the
magnitude of the thermoelectric effect detected in these experiments exceeded theoretical
estimates by {\it few orders of magnitude}. 

Later on large thermoelectric signals were also observed in multi-terminal hybrid superconducting-normal-superconducting (SNS) structures \cite{Venkat1,Venkat2,Petrashov03,Venkat3,Vitya} frequently called Andreev interferometers. Furthermore, the thermopower detected in these experiments was found to be periodic in the superconducting phase difference $\chi$ across the corresponding SNS junction. The latter observation (i) indicates that the thermoelectric signal in superconductors can be {\it phase coherent} and (ii) calls for establishing a clear relation between thermoelectric, Josephson and Aharonov-Bohm effects in systems under consideration. Both issues (i) and (ii) -- along with an experimentally observed large magnitude of the thermoelectric effect -- constitute the key subjects of our present review.

It is obvious from the above considerations that large thermoelectric effects (not restricted by a small parameter $T/\varepsilon_F$) in superconducting structures can be expected provided electron-hole symmetry in such structures is violated in some way. In this case the contributions from electron-like and hole-like excitations to the thermoelectric coefficient would not cancel each other anymore and, hence, $\alpha_S$ can become large. 

\begin{figure}
\centering
\includegraphics[width=60mm]{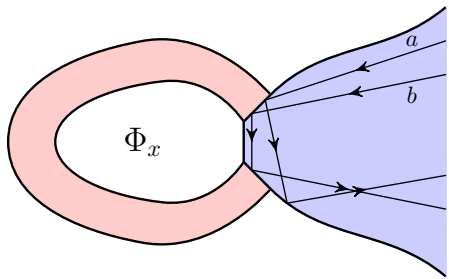}
\caption{A simple setup illustrating electron-hole symmetry breaking due to Andreev reflections (trajectory $b$). The setup consists of   a superconducting ring pierced by external magnetic flux $\Phi_x$ and attached to a piece of a normal metal. }
\label{nsnsn-fig}
\end{figure}

Electron-hole asymmetry in superconducting hybrid structures can be realized in a variety of physical situations. As a simple example, let us consider a superconducting ring pierced by external magnetic flux $\Phi_x$ and interrupted by a normal metal as it is schematically shown in Fig. \ref{nsnsn-fig}. Quasiparticles propagating from a normal metal towards a superconducting ring eventually hit either one NS interface (e.g., trajectory $a$) or both of them (e.g., trajectory $b$). In either case an incoming electron with energy $\varepsilon$ may be reflected back as a hole \cite{And}. In the low energy limit  $\varepsilon \ll \Delta$ and in the case of trajectory $a$ the probability for this Andreev reflection process  reads
\begin{equation}
\mathcal{R}_a^{e-h}(\varepsilon )=D_1^2/(1+R_1)^2,
\label{BTK}
\end{equation}
which is nothing but the standard BTK result \cite{BTK}. Here and below $D_{1,2}=1-R_{1,2}$ denote normal transmissions of the corresponding NS interfaces. 

Likewise, the Andreev reflection probability for electrons following trajectory $b$ can be derived in the form \cite{KZ17}
\begin{multline}
\mathcal{R}_{b}^{e-h} (\varepsilon )=
1-\frac{16R_1R_2}
{\bigl|
(1+R_1)(1+R_2)
+D_1D_2e^{i\left(\chi + \frac{2\varepsilon d}{v_F}\right)}
\bigr|^{2}},
\label{as2}
\end{multline}
where $d \gg v_F/\Delta$ is the effective distance covered by a quasiparticle between successive
scattering events at two NS interfaces and $\chi =2\pi \Phi_x/\Phi_0$ is the superconducting phase difference across our SNS junction.
In Eq. \eqref{as2} we again assume $\varepsilon \ll \Delta$.

Making use of general symmetry relations it is straightforward to verify
that the probability for an incoming hole to be reflected back as an
electron equals to $\mathcal{R}^{h-e}(\varepsilon)=\mathcal{R}^{e-h}(-\varepsilon)$. 
Combining this relation with Eq. \eqref{as2} we obtain 
\begin{equation}
\mathcal{R}_{b}^{e-h}(\varepsilon)\neq \mathcal{R}_{b}^{h-e}(\varepsilon ),
\label{pras}
\end{equation} 
implying that scattering on two NS interfaces generates {\it electron-hole symmetry violation} 
in our hybrid structure provided $\chi \neq \pi n$ and $0<D_{1,2}<1$. Below we will demonstrate that this
electron-hole asymmetry yields a large thermoelectric effect in the system
under consideration.

\section{Quasiclassical formalism and circuit theory}
\label{sec:quas}

\subsection{Eilenberger and Usadel equations}
\label{subsec:eilen}
Throughout our paper we will consider superconducting hybrid structures which can be
described by means of the standard quasiclassical formalism employing 
non-equilibrium Green-Keldysh matrix functions
$\check g$ which obey the Eilenberger equations \cite{bel}
\begin{equation}
-i\bm{v}_F\nabla \check g(\bm{p}_F,\bm{r},\varepsilon,t) =
[\check \Omega - \check \Sigma, \check g(\bm{p}_F,\bm{r},\varepsilon,t)],
\
\check g^2 =1.
\label{eilen}
\end{equation}
The check symbol denotes $4\times 4$ Keldysh matrices
\begin{equation}
\check X =
\begin{pmatrix}
\hat X^R & \hat X^K \\
0 & \hat X^A\\
\end{pmatrix},
\quad
X=g,\,\Omega ,\,\Sigma,
\end{equation}
with blocks $\hat X^{R,A,K}$ being $2\times2$ matrices in the Nambu space.
The matrix $\check \Omega$ has the standard structure
\begin{equation}
\hat \Omega^R = \hat \Omega^A =
\begin{pmatrix}
\varepsilon & \Delta \\
-\Delta  & -\varepsilon\\
\end{pmatrix},
\quad
\hat \Omega^K=0,
\end{equation}
where $\varepsilon$ is the quasiparticle energy, $\Delta$ is the superconducting order
parameter which is chosen real further below.

Scattering of electrons on impurities is
accounted for by the self-energy matrix $\check \Sigma$ which can be expressed
in the form
\begin{equation}
\check \Sigma = - i\Gamma \left<\check g\right> + \check \Sigma_{\mathrm{m}},
\quad
\Gamma=v_F/(2\ell).
\label{selfe}
\end{equation}
Here the first term describes the effect of non-magnetic isotropic
impurities while the second term  $\check \Sigma_{\mathrm{m}}$
is responsible for electron scattering on magnetic
impurities to be specified below.
Averaging over the Fermi surface is denoted by angular brackets
$\left<\cdots\right>$. 

Finally, the current density $\bm{j}$ is defined
with the aid of the standard relation
\begin{equation}
\bm{j}(\bm{r}, t)= -\dfrac{e N_0}{4} \int d \varepsilon
\left< \bm{v}_F \mathrm{Sp} [\hat \tau_3 \hat g^K(\bm{p}_F,
\bm{r},\varepsilon, t)] \right>,
\label{currenteilen}
\end{equation}
where $N_0$ is the electron density of states per spin direction at the
Fermi level and $\hat \tau_3$ is the Pauli matrix.

Provided elastic mean free path $\ell$ is small enough, i.e. $\ell \ll v_F/T_C$, Eilenberger equations \eqref{eilen} can be transformed into much simpler diffusion-like Usadel equations
\begin{gather}
\label{Usadel}
i D \nabla \left(\check G  \nabla \check G\right)=
\left[\check \Omega, \check G \right], \quad \check G \check G=1,
\end{gather}
where $\check G = \left<\check g(\bm{p}_F,\bm{r},\varepsilon,t)\right>$ is isotropic part of the Eilenberger Green function and $D = v_F \ell/3$ is the diffusion coefficient. The current density $\bm{j}$ in diffusion limit is related to the matrix $\check G$ in the standard manner as
\begin{gather}
\bm{j} =  -\dfrac{\sigma}{8 e }
\int
d \varepsilon
\Sp (\hat \tau_3\check G \nabla \check G)^K,
\end{gather}
where $\sigma = 2 e^2 N_0 D $ is the normal state Drude conductivity.

\subsection{Circuit theory for quasi-one-dimensional conductors}
\label{subsec:circuit}
In the diffusive limit the above quasiclassical theory of superconductivity can also be reformulated in the form that in some cases can be more convenient both for quantitative calculations and for qualitative analysis of the results. This our approach \cite{KZ2021PRB} essentially extends Nazarov’s circuit theory \cite{Naz1,Naz2} which can in principle be employed for conductors of arbitrary dimensionality. Here we focus our attention specifically on quasi-one-dimensional conductors where substantial simplifications can be achieved.

Let us first express Keldysh component $\hat G^K$ of the Green function matrix via retarded ($\hat G^R$) and advanced ($\hat G^A$) components of this matrix as
\begin{equation}
\hat G^K = \hat G^R \hat h - \hat h \hat G^A,
\label{GKRA}
\end{equation}
where $\hat h = h^L + \hat \tau_3 h^T$ is the matrix distribution function parameterized by two different quasiparticle distribution functions $h^L$ and $h^T$. In normal conductors the latter functions obey diffusion-like equations
\begin{gather}
D\nabla\left[ D^T\nabla h^T + \mathcal{Y} \nabla h^L + \bm{j}_{\varepsilon} h^L \right]=0,
\label{htusadel}
\\
D\nabla\left[ D^L\nabla h^L - \mathcal{Y} \nabla h^T + \bm{j}_{\varepsilon} h^T \right]=0,
\label{hlusadel}
\end{gather}
where $D^{T,L}$ and $\mathcal{Y}$ denote dimensionless kinetic coefficients and $\bm{j}_{\varepsilon}$ represents the spectral current
\begin{gather}
D^T 
=
\nu^2 + \dfrac{1}{4} |F^R + F^A|^2, 
\\
D^L 
=\nu^2 - \dfrac{1}{4} |F^R - F^A|^2, 
\\
\mathcal{Y} 
=-\dfrac{1}{4}\left(|F^R|^2 - |\tilde F^R|^2 \right), 
\\
\bm{j}_{\varepsilon} = 
\dfrac{1}{2}\Rea\Bigl( F^R \nabla \tilde F^R - \tilde F^R \nabla F^R \Bigr),
\label{jE}
\end{gather}
$\nu = \Rea G^R$ is the local density of states and $G^{R,A}$, $F^{R,A}$ and  $\tilde F^{R,A}$ are components of retarded and advanced Green functions
\begin{equation}
\hat G^{R,A}
=
\begin{pmatrix}
G^{R,A} & F^{R,A} \\
\tilde F^{R,A} & - G^{R,A}
\end{pmatrix}.
\end{equation}

In the case of quasi-one-dimensional conductors Eqs. \eqref{htusadel} and \eqref{hlusadel} are linear differential equations for the distribution functions $h^T$ and $h^L$ implying that there exist linear relations between these functions at different points $x$ and $\tilde x$. These relations can be conveniently represented in the Kirchhoff-like form \cite{KZ2021PRB}
\begin{equation}
\hat G_{x,\tilde x} 
\begin{pmatrix}
h^T(x) \\ h^L(x)
\end{pmatrix}
+ 
\hat G_{\tilde x,x}  
\begin{pmatrix}
h^T(\tilde x) \\ h^L(\tilde x)
\end{pmatrix}
=
- e 
\begin{pmatrix}
I^T \\ I^L
\end{pmatrix},
\label{kirmat}
\end{equation}
where 
\begin{equation}
\begin{pmatrix}
I^T \\ I^L
\end{pmatrix}
=-\dfrac{\mathcal{A} \sigma}{4 e}
\begin{pmatrix}
\Sp (\check G \check G' \hat \tau_3)^K \\ \Sp (\check G \check G')^K
\end{pmatrix}
\end{equation}
is the matrix current which remains conserved along the normal wire segment, $\hat G_{x,\tilde x}$ is $2\times 2$ conductance matrix which is in general a complicated function of the parameters $D^{T,L}$, $\mathcal{Y}$ and $\bm{j}_{\varepsilon}$. It obeys the following relations
\begin{equation}
\hat G_{x,\tilde x} + \hat \tau_3  \hat G_{\tilde x,x}^T \hat \tau_3 =0,
\quad
\hat G_{x,\tilde x} + \hat G_{\tilde x,x}
=
 \hat G_j,
\end{equation}
where we defined $\hat G_j = \hat\tau_1 G_j$ and $G_j = \sigma j_{\varepsilon} \mathcal{A}$.

The total electric current $I$ flowing across the wire is linked to the $I^T$-component of the matrix current by means of the equation
\begin{gather}
I= \frac{1}{2} \int I^T  d \varepsilon .
\label{currentI}
\end{gather}
In the normal wires connected to normal terminals we have $j_{\varepsilon} \equiv 0$ and matrix conductance can be evaluated explicitly
\begin{equation}
\hat G_{x,\tilde x}
=
\mathcal{A} \sigma
\left[\int_{\tilde x}^{x} dx'
\begin{pmatrix}
D^T(x') & \mathcal{Y}(x') \\
-\mathcal{Y}(x') & D^L(x')
\end{pmatrix}^{-1}
\right]^{-1}.
\label{GN}
\end{equation}
Under the conditions $|D^{T,L} - 1|\ll 1$ and $|\mathcal{Y}| \ll 1$ the above expression reduces to a simpler form
\begin{equation}
\hat G_{x,\tilde x}
\approx
\dfrac{\mathcal{A} \sigma}{(x - \tilde x)^2}
\int_{\tilde x}^{x} dx'
\begin{pmatrix}
D^T(x') & \mathcal{Y}(x') \\
-\mathcal{Y}(x') & D^L(x')
\end{pmatrix}.
\label{GNapprox}
\end{equation}

Further simplifications occur provided the energy of electrons propagating in the normal metal remains smaller than the superconducting order parameter $\Delta$ of the electrodes. At such energies the spectral current $I^L$ equals to zero inside normal wires connected to  superconducting terminals and we have
\begin{equation}
\hat G_{0,x}
=
\begin{pmatrix}
-G_S & 0 \\
G_j & 0
\end{pmatrix}
, 
\quad
\hat G_{x,0}
=
\begin{pmatrix}
G_S & G_j \\
0 & 0
\end{pmatrix},
\label{hatGSN}
\end{equation}
where it is assumed that an SN interface is located at $x=0$, $x$ is the coordinate inside the normal wire and $G_S(\varepsilon)$ is the spectral parameter characterizing both the interface and the N-wire. 

In general the matrix conductance of a quasi-one-dimensional wire can be parametrized as
\begin{gather}
\hat G_{x,\tilde x}
=
-
\begin{pmatrix}
G^T & G^{\mathcal{Y}} - G_j/2
\\
- G^{\mathcal{Y}} - G_j/2 & G^L
\end{pmatrix}.
\end{gather}
Note that diagonal elements of the matrix conductances ($G_S$, $G^{T,L}$) are even functions of both energy and the superconducting phase difference whereas their off-diagonal elements ($G^{\mathcal{Y}}$, $G_J$) are odd function of these two variables.

\section{Thermoelectric effect and spin-sensitive electron scattering}
\label{sec:thermo:sf}

In this section we will outline several examples demonstrating that spin-sensitive electron scattering 
in superconductors may generate electron-hole symmetry breaking which in turn 
yields dramatic enhancement of the thermoelectric effect. Our analysis is mainly based on Refs. 
\cite{KZK12,KZ14,KZ15M,KZ15}.

\subsection{Randomly distributed magnetic impurities}
\label{subsec:magimp}
Let us first consider a superconductor doped with randomly distributed point-like magnetic impurities with concentration $n_{\mathrm{imp}}$.
Electron scattering on such impurities can be described by the contribution $\check \Sigma_{\mathrm{m}}$ to the self-energy \eqref{selfe}
which takes the following general form \cite{Rusinov69}
\begin{multline}
\check \Sigma_{\mathrm{m}}=\dfrac{n_{\mathrm{imp}}}{2\pi N_0}
\Bigl\{
\left( [u_1 + \hat \tau_3 u_2]^{-1} +
i \left< \check g \right> \right)^{-1}
+\\+
\left( [u_1 - \hat \tau_3 u_2]^{-1} + i \left< \check g \right> \right)^{-1}
\Bigr\},
\label{selfem}
\end{multline}
where $u_{1,2}$ are dimensionless parameters characterizing the impurity scattering potential. 

It is worth pointing out that within the Born approximation the self-energy \eqref{selfem} 
reduces to the standard Abrikosov-Gor'kov result \cite{Abrikosov60}. The latter approximation is, however, insufficient for our purposes since it does not allow to capture the effect of Andreev bound states with energies
\begin{equation}
\varepsilon_B=\pm\beta\Delta ,\quad \beta^2 = \dfrac{( 1 + u_1^2-u_2^2 )^2}{( 1 + u_1^2-u_2^2 )^2 + 4 u_2^2},
\end{equation}
which are formed and localized near each paramagnetic impurity in a superconductor \cite{Shiba68,Rusinov68}. The presence of 
such bound states is crucial for electron-hole asymmetry in our system. Hence, within our further analysis it is necessary to go beyond the frequently employed Born approximation and make use of the self-energy in the form \eqref{selfem}.

In order to proceed we apply a temperature gradient $\nabla T$ to our superconductor and resolve Eqs. \eqref{eilen} together with \eqref{selfe} and \eqref{selfem} evaluating the linear correction to the Green-Keldysh function
$\delta \hat g^K \propto \bm{v}_F\nabla T$. It is straightforward to verify that charge neutrality is explicitly maintained inside the superconductor,
since the induced voltage vanishes identically, $V \propto \langle \Sp \delta \hat g^K\rangle \propto \langle \bm{v}_F\rangle \equiv 0$.
The whole calculation goes along the lines with the analysis of thermal conductivity in unconventional superconductors \cite{Graf96}.   
Combining the resulting expression for $\delta \hat g^K$ with Eq. \eqref{currenteilen} we recover
the dominating contribution to the thermoelectric coefficient $\alpha_S$ which originates from electron-hole
asymmetry. It reads \cite{KZK12}
\begin{gather}
\alpha_S=-\dfrac{eN_0 v_F^2}{12 T^2}
\int_{-\infty}^{\infty}
\dfrac{\mathcal{F}(\varepsilon)d\varepsilon}{\cosh^2(\varepsilon/2T)},
\label{thermo:koeff}
\\
\mathcal{F}(\varepsilon)=\dfrac{ \varepsilon \nu(\varepsilon)\Img\Sigma_0^R(\varepsilon)}{
\left[\Img\mathcal{D}(\varepsilon)\right]^2 -
\left[\Img\Sigma_0^R(\varepsilon)\right]^2}.
\label{calF}
\end{gather}
Here $\nu(\varepsilon)$ defines  the energy resolved superconducting density of
states in the presence of magnetic impurities normalized to its normal state value and $\Sigma_0^R$ is
a non-vanishing diagonal part of the retarded self-energy matrix in Eqs. \eqref{selfe} and \eqref{selfem} which explicitly accounts for asymmetry between electrons and holes. We obtain
\begin{gather}
\Sigma_0^R(\varepsilon)=
\Gamma_0
\dfrac{\tilde \varepsilon^2 - \strut \Delta^2 }{
\tilde \varepsilon^2 - \beta^2 \strut \Delta^2 }, \quad \nu(\varepsilon)=\Rea\dfrac{\tilde \varepsilon}{\sqrt{\tilde \varepsilon^2 - \Delta^2}},
\label{nusigma0}
\\
\mathcal{D}(\varepsilon)=
\sqrt{\tilde \varepsilon^2 - \Delta^2} + i\Gamma +
i \Gamma_1
\dfrac{\tilde \varepsilon^2 - \Delta^2}{
\tilde \varepsilon^2 - \beta^2 \Delta^2 },
\label{sqrt}
\end{gather}
where the parameter $\tilde \varepsilon$ is fixed
by the relation \cite{Shiba68,Rusinov68}
\begin{equation}
\tilde \varepsilon = \varepsilon
+i\Gamma_2
\dfrac{\tilde \varepsilon \sqrt{\tilde \varepsilon^2 - \Delta^2}
}{\tilde \varepsilon^2 - \beta^2 \Delta^2},
\end{equation}
The scattering parameters $\Gamma_{0,1,2}$ are defined as
\begin{gather}
\Gamma_0=\dfrac{n_{\mathrm{imp}}}{\pi N_0}\dfrac{u_1 ( 1 + u_1^2-u_2^2 )}{( 1 + u_1^2-u_2^2 )^2 + 4 u_2^2  },
\label{Gamma0}
\\
\Gamma_1=
\dfrac{n_{\mathrm{imp}}}{\pi N_0}
\dfrac{( 1 + u_1^2-u_2^2 ) (u_1^2-u_2^2)}{( 1 + u_1^2-u_2^2 )^2 + 4 u_2^2  }
\label{Gamma1}
\\
\Gamma_2 =2\dfrac{n_{\mathrm{imp}}}{\pi N_0}\dfrac{u_2^2}{( 1 + u_1^2-u_2^2 )^2 + 4 u_2^2  }.
\label{Gamma2}
\end{gather}
Note that $\tilde \varepsilon$, $\Sigma_0^R$ and $\nu(\varepsilon)$ do not depend $\Gamma$, i.e. all these parameters remain insensitive to 
electron scattering on non-magnetic impurities because such kind of scattering does not produce any pair-breaking effect
in conventional superconductors. On the contrary, scattering on magnetic impurities may strongly modify these parameters.

\begin{figure}
\centering
\includegraphics[width=80mm]{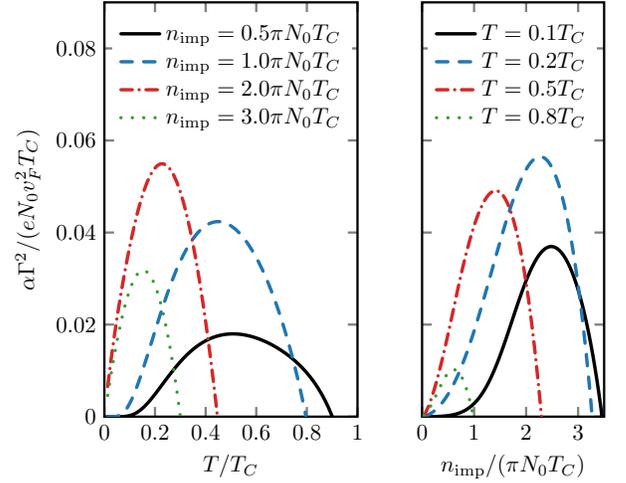}
\caption{Thermoelectric coefficient as a function of temperature and magnetic impurity concentration.
Scattering parameters $u_1=u_2=0.5$ and the scattering rate $\Gamma=10T_{C}$ remain the same for both panels. }
\label{tp-g10-fig}
\end{figure}

In the most relevant case of diffusive
superconductors with $\Gamma \gtrsim T_C$ Eq. \eqref{calF} reduces to
$\mathcal{F}(\varepsilon)=\nu(\varepsilon)\Img\Sigma_0^R(\varepsilon)/\Gamma^2$, i.e. $\alpha_S\propto 1/\Gamma^2$ in this limit. 
Further limiting expressions for $\alpha_S$ are summarized elsewhere \cite{KZK12}. Here we only provide the results of numerical evaluation of $\alpha_S$ as a function of both temperature and impurity concentration. These results are displayed in Fig. \ref{tp-g10-fig}. 

We observe that the thermoelectric coefficient for a diffusive superconductor
achieves its maximum value at $T\sim T_C/2$ and $n_{\mathrm{imp}}$ approximately equal to one-half of the critical concentration
at which superconductivity gets fully suppressed. This maximum value takes the form
\begin{equation}
\max_{T, n_{\mathrm{imp}}} |\alpha_S| \approx 0.05 \dfrac{e N_0 v_F^2 T_{C}}{\Gamma^2}=
0.2 e N_0 T_{C} \ell^2 .
\label{alphaSopt}
\end{equation}
Combining the relation $\alpha_N \sim (\sigma_N/e)(T/\varepsilon_F)$ with Eq. \eqref{alphaSopt} we arrive at a simple estimate \cite{KZK12}
\begin{equation}
\alpha_S/\alpha_N(T_C) \sim p_F \ell \gg 1,
\label{est}
\end{equation}
demonstrating that the enhancement of the thermoelectric effect is stronger in cleaner superconductors. At the
borderline of applicability of Eq. \eqref{est} $\ell \sim v_F/T_C$ we obtain
$|\alpha_S |\sim \sigma_N/e$, which defines the absolute maximum value of  $\alpha_S$ in conventional superconductors doped by magnetic impurities.

\begin{figure}
\centering
\includegraphics[width=50mm]{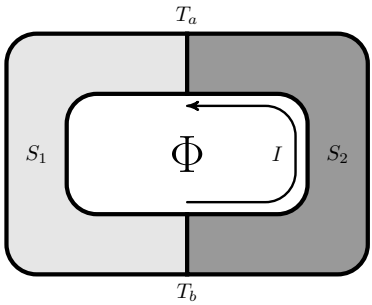}
\caption{A ring formed by two different superconductors with contacts
maintained at different temperatures $T_a$ and $T_b$.}
\label{ss-fig}
\end{figure}

As we already mentioned in Sec. \ref{sec:intro}, one possible way to detect the thermoelectric in superconductors is to perform an experiment with bimetallic superconducting rings \cite{Zavaritskii74,Falco76,Harlingen80} as it is schematically shown in Fig. \ref{ss-fig}. Provided
superconducting contacts are kept at different temperatures $T_a$ and $T_b$, thermoelectric current is
generated inside the ring and the corresponding magnetic flux $\Phi$ can be detected. The magnitude of this flux is defined as
\begin{equation}
\dfrac{\Phi }{\Phi_0}=\dfrac{4e}{c^2}
\int_{T_a}^{T_b}
[\lambda_1^2(T)\alpha_{S1}(T)-\lambda_2^2(T)\alpha_{S2}(T)]dT,
\label{phi-thick}
\end{equation}
where $\Phi_0 = \pi c/e$ is flux quantum, $\alpha_{S1,2}$ and $\lambda_{1,2}$ denote respectively thermoelectric coefficients and
the values of London penetration depth for two superconductors. For the sake simplicity we may assume $\alpha_{S1}\gg \alpha_{S2}$ and neglect the second term in Eq. \eqref{phi-thick}. Employing Eq. \eqref{alphaSopt} together with the standard expression for the London penetration depth
in diffusive superconductors at $T=0$ we obtain a conservative estimate for the thermally induced flux
\begin{equation}
\dfrac{|\Phi |}{\Phi_0} \sim 0.01 \dfrac{|T_b - T_a|}{\Gamma}, \quad \Gamma \gtrsim T_{C},
\end{equation}
implying that for reasonably clean superconductors typical values of $\Phi$ may easily reach $\Phi\gtrsim 10^{-2}\Phi_0$.

\subsection{Spin-active interfaces}
\label{subsec:spinactive}
\begin{figure}
\centering
\includegraphics[width=70mm]{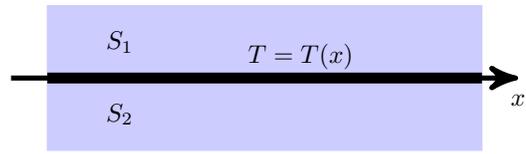}
\caption{A metallic bilayer which consists of two superconductors $S_1$ and $S_2$ separated by a spin-active interface. The temperature $T(x)$ changes only in the direction parallel to the interface.}
\label{sfs-fig}
\end{figure}

As another example of a system with intrinsically induced electron-hole asymmetry let us consider an extended bilayer consisting of two metallic slabs, one  superconducting and another one being either normal or superconducting as well. This structure is schematically displayed in Fig. \ref{sfs-fig}. In what follows we will assume that both metals are brought into direct contact with each other via a spin-active interface that is located in the plane $z=0$. Such an interface can be formed, e.g., by an ultra-thin layer of a ferromagnet.

A minimal (though sufficient) model for our spin-active interface involves three
different parameters, i.e. the transmission probabilities for opposite spin directions
$D_{\uparrow}$ and $D_{\downarrow}$ as well as the so-called
spin mixing angle $\theta$ representing the difference between the scattering
phase shifts for spin-up and spin-down electrons. These parameters are assumed to be energy 
independent which can be justified for sufficiently thin ferromagnetic layers. At the same time, the layer should not be too thin
in order to preserve its ferromagnetic state.

Finally, we will assume that the left ($x \to -\infty$) and right ($x \to \infty$) edges of our bilayer are maintained at temperatures $T_a$ and $T_b$ respectively (see Fig. \ref{sfs-fig}). Hence, quasiparticles entering our system from the left (right) side are described by the equilibrium (Fermi) distribution function with temperature $T_a$ ($T_b$). For the sake of simplicity we will also assume that temperature $T$ in our bilayer structure depends only on $x$ and does not vary along $y$- and $z$-directions.

\begin{figure}
\centering
\includegraphics[width=70mm]{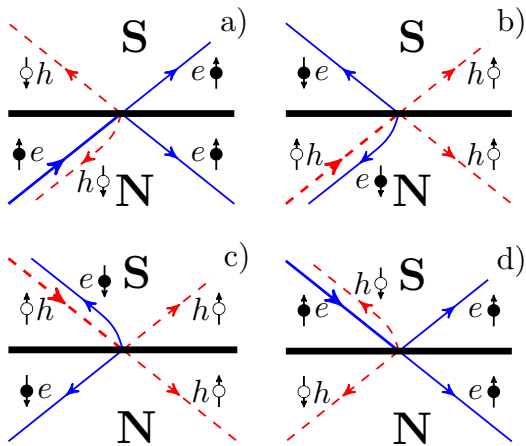}
\caption{Four different electron and hole scattering processes in a
superconducting-normal bilayer.}
\label{ehscat}
\end{figure}

For the sake of definiteness let us address an SFN structure and distinguish 16 different scattering processes illustrated
in Fig. \ref{ehscat}. Depending on whether incident electron-like or hole-like excitations arrive from the normal metal
or the superconductor one can classify all these processes into four groups (a), (b), (c) and (d). Consider, for instance, the four scattering processes of an electron-like excitation propagating towards the NS interface from the normal metal side, see Fig. \ref{ehscat}a. Provided the energy of this excitation $\varepsilon$ does not exceed $\Delta$, it cannot penetrate deep into the superconductor and gets reflected
back into the normal metal either as an electron (specular reflection) with probability $\mathcal{R}^{e-e}_{NS,\sigma}$
or as a hole (Andreev reflection) with probability $\mathcal{R}^{e-h}_{NS,\sigma}$. The probabilities for all remaining scattering processes are defined analogously.

In the case of spin-independent scattering $R_{\uparrow}=R_{\downarrow}$ and $\theta=0$ both transmission and reflection probabilities remain symmetric under the replacement of an electron by a hole and vice versa, i.e. we have, e.g., $\mathcal{R}^{e-e}_{NS,\sigma} = \mathcal{R}^{h-h}_{NS,\sigma}$, $\mathcal{R}^{e-h}_{NS,\sigma} = \mathcal{R}^{h-e}_{NS,\sigma}$ and so on. In this case we just recover the standard BTK results \cite{BTK} (cf., e.g., Eq. \eqref{BTK}) and no violation of electron-hole symmetry can be expected. By contrast, in the case of spin-sensitive scattering with $R_{\uparrow} \neq R_{\downarrow}$ and $\theta \neq 0$ it is straightforward to demonstrate \cite{KZ14} that only two reflection probabilities remain equal,
$\mathcal{R}^{e-h}_{NS,\sigma}=\mathcal{R}^{h-e}_{NS,\sigma}$, whereas all others differ, e.g., $\mathcal{R}^{e-e}_{NS,+} \neq \mathcal{R}^{e-e}_{NS,-}$,
$\mathcal{R}^{e-e}_{NS,\sigma} \neq \mathcal{R}^{h-h}_{NS,\sigma}$, $\mathcal{R}^{e-h}_{NS,+}\neq\mathcal{R}^{h-e}_{NS,-}$, etc. Thus, we may conclude that electron scattering at spin-active interfaces generates electron-hole imbalance in SFN structures which manifests itself by different scattering rates for electrons and holes in such systems. A similar conclusion can also be reached in the case of bilayer structures formed by two superconductors \cite{KZ15M}.

Note that to a certain extent this situation resembles the one already addressed in Sec. \ref{sec:intro}, cf. Eq. \eqref{pras}. However, despite some qualitative similarities the physical origin for this effect in the latter case is, of course, not exactly the same. In some sense, the physical picture considered here is closer to that addressed above in Sec. \ref{subsec:magimp} where spin-sensitive electron on local magnetic impurities was analyzed. In this respect spin-active interfaces can just be treated as spatially extended magnetic impurities.

Now let us evaluate an electric current response to a temperature gradient applied to the system along the metallic interface. 
In order to proceed we need to again resolve quasiclassical Eilenberger equations \eqref{eilen} supplemented
by proper boundary conditions describing electron transfer across SFS-interfaces by matching the 
Green function matrices $\check g$ for incoming and outgoing momentum directions
at both sides of this interface. In the case of spin-active interfaces the corresponding boundary conditions
have been derived in Ref. \cite{Millis88}. An equivalent approach has been worked out in Ref. \cite{Zhao04}.

In a general case $D_{\uparrow}\neq D_{\downarrow}$, $\theta \neq 0$ and provided two superconductors
are not identical (one of them can also be a normal metal) the thermoelectric current can be evaluated numerically or analytically
in certain limits \cite{KZ14,KZ15M,KZ15}. In the presence of particle-hole asymmetry this current can be large and strongly exceed  its values, e.g., in normal metals. Here we will not go into details of the calculation and only present simple order-of-magnitude estimates. For ballistic bilayers the magnitude of the thermoelectric current density can roughly be estimated as  \cite{KZ14,KZ15M}
\begin{equation}
|\bm{j}(z)| \sim j_c (R_{\uparrow} - R_{\downarrow}) \sin\theta \frac{T_a - T_b}{T_C},
\label{jT}
\end{equation}
where $j_c\sim e v_F N_0T_C$ is the critical current of a clean superconductor
with the critical temperature $T_C$. We observe that, in contrast to the situation considered in \cite{Galperin73,Aronov81}, Eq. \eqref{jT}
does not contain the small factor $T/\varepsilon_F \ll 1$, i.e. the thermoelectric effect becomes really large. For instance, by setting $(R_{\uparrow} - R_{\downarrow}) \sin\theta \sim 1$ and $T_1 - T_2 \sim T_C$, one achieves the thermoelectric
current densities of the same order as the critical one $j_c$.

In the tunneling limit $D_{\uparrow}, D_{\downarrow} \ll 1$ and for the case of diffusive superconductors with very different mean free path values (i.e. for $\ell_1^2 + \ell_2^2 \gg \ell_1
  \ell_2$) the expression for the thermoelectric current takes a simple form \cite{KZ15}
\begin{multline}
I=
\dfrac{e N_0}{8v_F^2} \partial_x T
\int
\dfrac{(\ell_1^2+\ell_2^2) \varepsilon  d \varepsilon}{T^2
\cosh^2(\varepsilon/2T)}
\Bigl<
v_{x}^2 |v_{z}|
\theta (-v_{z})
\times\\\times
(D_{\uparrow} - D_{\downarrow})
\left[\nu_{1\uparrow}(0) \nu_{2\uparrow}(0)-
\nu_{1\downarrow}(0) \nu_{2\downarrow}(0) \right]
\Bigr>,
\label{thermoIdiff2}
\end{multline}
where $\nu_{i\uparrow(\downarrow)}(0)$ are the momentum and energy resolved densities
of states at the interface for the opposite electron spin orientations. At intermediate temperatures 
$T \sim \Delta$ we can roughly estimate the magnitude of the thermoelectric current as
\begin{equation}
I \sim e N_0 v_F\ell^2 (R_{\uparrow} - R_{\downarrow}) \sin\theta\partial_x T .
\label{estim}
\end{equation}
It is instructive to compare this result with that for the thermoelectric current $I_{\text{norm}}$ flowing in our bilayer in its normal state. Making use of the well known Mott relation for the thermoelectric coefficient of normal metals, from \eqref{estim} we obtain
\begin{equation}
\dfrac{I}{I_{\text{norm}}}
\sim
\dfrac{\ell}{d}\dfrac{\varepsilon_F}{T_C}(R_{\uparrow} - R_{\downarrow})\sin\theta,
\label{est2}
\end{equation}
where $d$ is the total thickness of our bilayer. Setting $\ell \sim d$, $R_{\uparrow} - R_{\downarrow} \sim 1$ and $\sin\theta \sim 1$ we immediately arrive at the conclusion that owing to electron-hole asymmetry the thermoelectric current $I$ in the superconducting state can be enhanced by a large factor up to $\varepsilon_F/T_C \gg 1$ as compared to that in the normal state $I_{\text{norm}}$.

\subsection{Related mechanisms and effects}
\label{subsec:rel}
The mechanisms of electron-hole symmetry breaking in superconducting hybrid structures are, of course, not limited to those considered above. Electron-hole asymmetry accompanied by large scale thermoelectric effects was also predicted to occur in superconductor-ferromagnet hybrids with the density of states spin split by the exchange and/or Zeeman fields \cite{Machon,Ozaeta}. 
These theoretical predictions were verified in experiments with superconductor-ferromagnet tunnel junctions in high magnetic
fields \cite{Beckmann} where large thermoelectric currents were observed. 
In unconventional superconductors electron-hole symmetry breaking and enhanced thermoelectric effects can also be caused by electron scattering on non-magnetic impurities \cite{Arfi88,Lofwander04}. In this case -- quite similarly to the situation considered in Sec. \ref{subsec:magimp} -- the crucial role is played by localized Andreev bound states formed inside the system.

Finally, we mention that electron-hole asymmetry in superconducting structures may also yield anomalously large photovoltaic effect \cite{Zaitsev86,KZ15a}. Despite some similarities the latter effect is rather different from the thermoelectric one analyzed here because thermal heating of the system is physically not equivalent to that produced by an external ac field. For more details on this issue we refer the reader to the work \cite{KZ15a}.

\section{Phase coherence and thermoelectricity in Andreev interferometers}
\label{sec:phase-coher}

In the previous section we demonstrated that electron-hole asymmetry leading to huge thermoelectric effects in superconducting structures can be generated by spin-sensitive scattering of quasiparticles on various magnetic inclusions. Large thermoelectric effects not restricted
by a small parameter $T/\varepsilon_F$ were also observed in the absence of such inclusions in multi-terminal hybrid superconducting structures frequently called Andreev interferometers  \cite{Venkat1,Venkat2,Petrashov03,Venkat3,Vitya}. Electron-hole symmetry breaking appears to be responsible for this phenomenon also in that case, and the key mechanism for such symmetry breaking (elucidated in Ref. \cite{KZ17} and also in the Introduction) is directly related to Andreev reflection at NS interfaces in the presence of magnetic flux, cf. Eqs. \eqref{as2} and \eqref{pras}.  

Furthermore, thermoelectric signals detected in experiments \cite{Venkat1,Venkat2,Petrashov03,Venkat3,Vitya}
demonstrated $2\pi$-periodic dependence on the superconducting phase difference $\chi$ across an SNS junction
forming an Andreev interferometer. The latter observation implies that in such structures thermoelectricity is phase-coherent. 
Earlier this issue was addressed by a number of authors \cite{Volkov,V2,V3,VH,VH2,VH3}.
It is also remarkable that that both odd and even in $\chi$ periodic dependencies
of the thermopower were observed in different interferometers. Several attempts to interpret this odd-even effect were performed attributing it, e.g., to charge imbalance \cite{Titov} and mesoscopic fluctuations \cite{JW}.

\begin{figure}
\centering
\includegraphics[width=70mm]{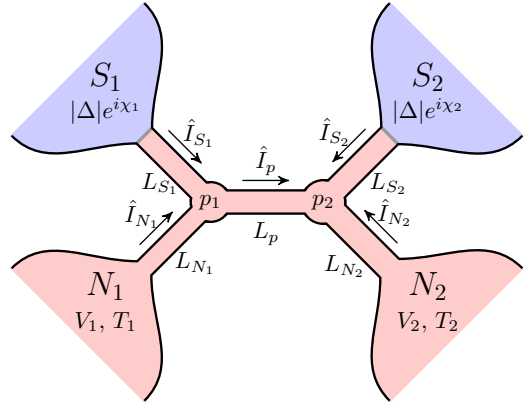}
\caption{Four terminal Andreev interferometer under consideration.}
\label{snns3-fig}
\end{figure}

In what follows we will consider Andreev interferometers schematically displayed in Fig. \ref{snns3-fig}. They consist of two normal and two superconducting terminals interconnected by five normal metallic wires of different lengths. Normal terminals are generally maintained at electrostatic potentials $V_1$ and $V_2$ and temperatures $T_1$ and $T_2$. The phase difference between two superconducting terminals equals to $\chi=\chi_2-\chi_1$.

The circuit theory formalism outlined in Sec. \ref{subsec:circuit} allows one to exactly solve the kinetic equation for the above structure expressing electric currents inside metallic wires via quasiparticle distribution functions in the terminals and spectral conductances of these wires. For instance, for symmetric four-terminal Andreev interferometers the currents flowing into normal ($I_N$) and superconducting ($I_S$) terminals read
\begin{gather}
I_N = - \dfrac{1}{4}
\int
G_N^{eff} 
(h_{N_1}^T - h_{N_2}^T)
d \varepsilon,
\label{INVbias0}
\\
I_S = -\dfrac{1}{2}
\int
\biggl[
G_j (h_{N_1}^L + h_{N_2}^L)
+
G_S^{eff} 
(h_{N_1}^T - h_{N_2}^T)
\biggr]
d \varepsilon,
\label{ISVbias0}
\end{gather}
where $G_N^{eff}$ and $G_S^{eff}$ are expressed in terms of the spectral conductances as
\begin{gather}
G_N^{eff}=
\dfrac{
(G_S + 2 G_p^T) [G_N^L G_N^T + (G_N^{\mathcal{Y}})^2]
}{G_N^L (G_N^T + G_S + 2 G_p^T) + (G_N^{\mathcal{Y}})^2
},
\label{GNeff0}
\\
G_S^{eff}
=
\dfrac{G_S [ G_N^T G_N^L + (G_N^{\mathcal{Y}})^2] - G_j G_N^{\mathcal{Y}} (G_S + 2 G_p^T)
}{G_N^L (G_N^T + G_S + 2 G_p^T) + (G_N^{\mathcal{Y}})^2}.
\label{GSeff0}
\end{gather}
Making use of the current conservation condition one also finds
\begin{equation}
\int 
\Bigl[
G^T_{eff}
(h_{N_1}^T + h_{N_2}^T)
+
G_{eff}^{\mathcal{Y}}
(h_{N_1}^L - h_{N_2}^L)
\Bigr]
d \varepsilon
=0,
\label{eqV2}
\end{equation}
where
\begin{gather}
G^T_{eff}
=
\dfrac{G_N^T (2 G_S G_p^L + G_j^2) + G_S [G_N^T G_N^L + (G_N^{\mathcal{Y}})^2]}{
(G_S + G_N^T) (2 G_p^L + G_N^L) + (G_j + G_N^{\mathcal{Y}})^2
},
\label{GTeff}
\\
G_{eff}^{\mathcal{Y}}=
\dfrac{ G_N^{\mathcal{Y}} (G_S 2 G_p^L + G_j^2) + G_j [G_N^T G_N^L + (G_N^{\mathcal{Y}})^2] 
}{
(G_S + G_N^T) (2 G_p^L + G_N^L) + (G_j + G_N^{\mathcal{Y}})^2
}.
\label{GYeff}
\end{gather}
Eqs. \eqref{INVbias0}-\eqref{GYeff} provide a formally exact solution describing the current distribution in symmetric four-terminal Andreev interferometers displayed in Fig.  \ref{snns3-fig}.

In what follows we will address two different types of external conditions which could be imposed on our structure. Specifically, normal terminals could be connected to an external battery that fixes the voltage drop $V=V_1 - V_2$ between these terminals. Alternatively, normal terminals could be totally disconnected from any external circuit in which case one obviously has $I_N \equiv 0$. In both cases the distribution functions $h^{T,L}$ in normal terminals have the form
\begin{equation}
h^{T/L}_{N_i}
=
\dfrac{1}{2}
\left[
\tanh\dfrac{\varepsilon + eV_i}{2T_i} \mp \tanh\dfrac{\varepsilon - eV_i}{2T_i}
\right],
\label{hTL}
\end{equation}
where $V_i$ and $T_i$ are voltage bias and temperature of the corresponding terminal.

\subsection{Interplay between Josephson and Aharonov-Bohm effects}
\label{subsec:ABJ}
To begin with, let us ignore thermoelectric effects by setting $T_1=T_2=T$ and fix the wire lengths and cross sections respectively as
$L_{S_1(N_1)}=L_{S_2(N_2)} = L_{S(N)}$ and $\mathcal{A}_{S_1(N_1)}=\mathcal{A}_{S_2(N_2)} = \mathcal{A}_{p}= \mathcal{A}$. In this case the current conservation condition \eqref{eqV2} obviously yields $V_{1/2} = \mp V/2$ and Eqs. \eqref{ISVbias0}, \eqref{INVbias0} can be rewritten in a simpler form observing the conditions $h^T_{N_1} = - h^T_{N_2}$ and $h^L_{N_1} = h^L_{N_2}$.

\begin{figure}
\centering
\includegraphics[width=80mm]{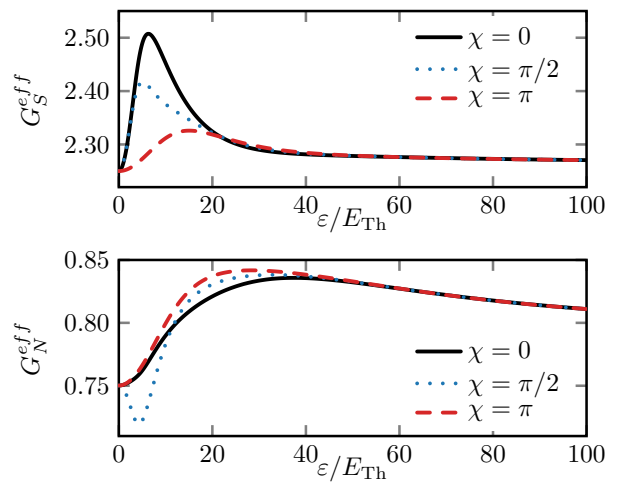}
\caption{Energy dependence of the spectral conductances for different values of $\chi$. Solid, dotted and dashed curves correspond to the phase difference $\chi$ equal to $0$, $\pi/2$ and $\pi$ respectively. Here we set $L_{S}=L_{N}=L_p$ and $E_{\mathrm{Th}} = 10^{-3}\Delta$. }
\label{GSNeff-fig}
\end{figure}

In Fig. \ref{GSNeff-fig} we show typical energy dependence of the spectral conductances $G_{S,N}^{eff}$ (which are even function of both energy and the superconducting phase difference) for different values of $\chi$. In the low and high energy limits these conductances approach the same value as a consequence of the well known reentrance effect \cite{bel}. 

It is convenient to represent the spectral conductances as a sum of three terms
\begin{equation}
G_{S,N}^{eff}(\varepsilon, \chi) = G_{S,N}^{eff,n} + 
\delta G_{S,N}^{eff}(\varepsilon) + \delta G_{S,N}^{AB}(\varepsilon, \chi),
\end{equation}
where $G_{S,N}^{eff,n}$ are the normal state values, $\delta G_{S,N}^{eff}(\varepsilon)$ represent the phase independent corrections due to the proximity effect and $\delta G_{S,N}^{AB}(\varepsilon, \chi)$ define periodic in $\chi$ Aharonov-Bohm-type corrections due to interference effects. The terms $\delta G_{S,N}^{eff}(\varepsilon)$ decay very slowly at high energies ($\sim |\varepsilon|^{-1/2}$) whereas $\delta G_{S,N}^{AB}(\varepsilon, \chi)$ decay exponentially on the energy scale of several Thouless energies ($\sim e^{-\sqrt{|\varepsilon|/E_{\mathrm{Th}}}}$). With this in mind we obtain $I_N \approx G_{S,N}^{eff,n} V$ with a small extra excess current $\sim V^{1/2}$ at large voltages and the phase-periodic component which amplitude does not depend on voltage in the high voltage limit.

Analogously, we can write the current $I_S$ flowing between superconducting terminals as a sum of three terms
\begin{equation}
I_S = I_0(V) + I_J(V,\chi)  + I_{AB}(V,\chi),
\label{Iphi}
\end{equation}
where $I_0$ is the phase independent contribution, while $I_J$  and $I_{AB}$ represent respectively odd and even functions of the phase difference with zero mean value. At zero voltage $I_{J}$ simply coincides with the equilibrium Josephson current \cite{ZZh,GreKa}. With increasing voltage the amplitude of the phase oscillations of $I_{J}(\chi )$ decreases whereas the amplitude of $I_{AB}(\chi)$ grows linearly $\propto V$ at low voltages and saturates at high voltages.

Equation \eqref{Iphi} can also be rewritten in the form
\begin{equation}
 I_S=I_0(V) + I_1(V, \chi - \chi_0(V)),
 \quad I_1(V, 0)=0.
\end{equation}
where we introduced the phase shift $\chi_0$ induced by the external voltage. This equation demonstrates that the net current $I_S(\chi )$ is in general no longer an odd function of $\chi$.

The physics behind this result is transparent. In the presence of a non-zero bias $V$ a dissipative current component, which we will further label as $I_d(V)$,
is induced in the normal wire segments $L_{S_1}$ and $L_{S_2}$. At NS interfaces this current gets converted into extra ($V$-dependent) supercurrent flowing across a superconducting loop.
Since at low temperatures and energies electrons in normal wires attached to a superconductor remain coherent (thus keeping information about the phase $\chi$), dissipative currents in such wires also become phase (or flux) dependent demonstrating even in $\chi$
Aharonov-Bohm-like (AB) oscillations \cite{nakano1991quasiparticle,SN96,GWZ97,Grenoble}, i.e.
$I_d(V,\chi)=I_0(V)+I_{AB}(V,\chi )$, where $I_0(V) \propto V$.
Combining this contribution to the current $I_S$ with an (odd in $\chi$) Josephson current $I_J(V,\chi)$ we immediately arrive at Eq. \eqref{Iphi}
with $I_1=I_J+I_{AB}$.

\begin{figure}
\centering
\includegraphics[width=70mm]{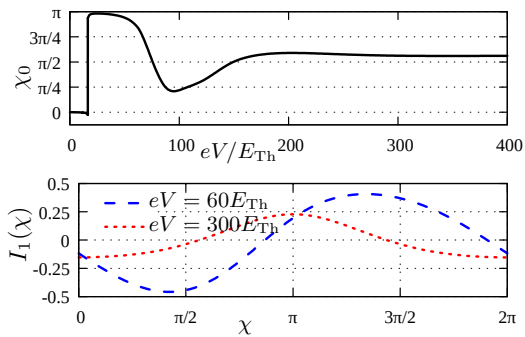}
\caption{Upper panel: The phase shift $\chi_0$ as a function of $V$. Lower panel: The current $I_1(\chi)$ at $ T \to 0$ for $eV = 60 E_{\mathrm{Th}}$ and $eV = 300 E_{\mathrm{Th}}$. Here we set $L_{S}=L_{N}=L_p$ and $E_{\mathrm{Th}} = 10^{-3}\Delta$. }
\label{fig: MainRes_VirtGeom}
\end{figure}

The behavior of the phase shift $\chi_0(V)$ displayed in Fig. \ref{fig: MainRes_VirtGeom} is determined by
a trade-off  between Josephson and Aharonov-Bohm contributions to $I_1$. At low voltages $I_J$ dominates over $I_{AB}$,
and one has $\chi_0 \approx 0$. Increasing the bias to values $eV \sim 20E_{\mathrm{Th}}$, in full agreement with the results \cite{WSZ} we observe the transition to a $\pi$-junction state meaning the sigh change of $I_J$. Here and below $E_{\mathrm{Th}}=D/L^2$ is the Thouless energy of our device and $L = 2L_S + L_p$ is the total length of three wire segments between two S-terminals. At even higher bias voltages both terms $I_J$ and $I_{AB}$ become of the same order. For $v=(eV/ 2 E_{\mathrm{Th}})^{1/2} \gg 1$ and at $T \ll E_{\mathrm{Th}}$
we have \cite{WSZ} $I_J =I_C(V)\sin\chi$, where for our geometry
\begin{equation}
I_C(V) \simeq \frac{128(1+v^{-1})}{9(3+2\sqrt{2})}\frac{V}{R_L} e^{ -v} \sin (v+v^{-1}).
\end{equation}
We also approximate $I_{AB} \approx I_{\mathrm{m}}\cos{\chi}$, where $I_{\mathrm{m}} \approx 0.18E_{\mathrm{Th}}/eR_L$ and $R_L$ is the normal resistance of the wire with length $L$. Then for $eV \gg E_{\mathrm{Th}} \gg T$ we get
\begin{equation*}
I_1 \approx \sqrt{I_C^2+I_{\mathrm{m}}^2}\sin (\chi-\chi_0), \quad \chi_0(V) = -\arctan\frac{I_{\mathrm{m}}}{I_C(V)}.
\end{equation*}
The function $\chi_0(V)$ demonstrates damped oscillations and saturates to the value $\chi =\pi/2$ in the limit of large voltages, as it is also illustrated in Fig. \ref{fig: MainRes_VirtGeom}.

At higher $T > E_{\mathrm{Th}}$ the Josephson current decays exponentially with increasing $T$ while the Aharonov-Bohm term
is described by a much weaker power-law dependence \cite{GWZ97,Grenoble} $I_{AB} \propto 1/T$, thus dominating the expression for $I_1$ and implying that $\chi_0 \simeq \pi/2$ at such values of $T$.

Finally, we note that the effects similar to those discussed here have recently been observed in non-magnetic Andreev interferometers \cite{Marg}.

\subsection{Thermopower}
\label{subsec:thermo}
\begin{figure}
\centering
\includegraphics[width=70mm]{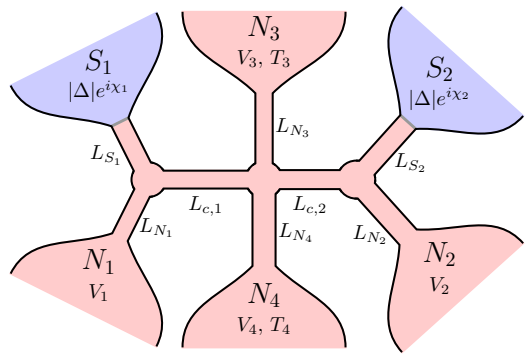}
\caption{Schematics of a six-terminal Andreev interferometer consisting of two superconducting and four normal terminals interconnected by normal metallic wires of different lengths.}
\label{fig: geom2}
\end{figure}

Let us now turn to the thermoelectric effect. For the sake of generality we slightly 
modify the setup in Fig. \ref{snns3-fig} by attaching two extra normal terminals N$_3$
and N$_4$ as shown in Fig. \ref{fig: geom2}. These terminals are disconnected
from the external circuit and are maintained at different
temperatures $T_3$ and $T_4$, while the temperature of the remaining
four terminals equals to $T$. 

Let us define the thermopower as
\begin{equation}
\mathcal{S}_{34} = \frac{V_3-V_4}{T_3-T_4}\Big|_{T_4 \mapsto T_3}.
\label{defS}
\end{equation}

\begin{figure}
\centering
\includegraphics[width=70mm]{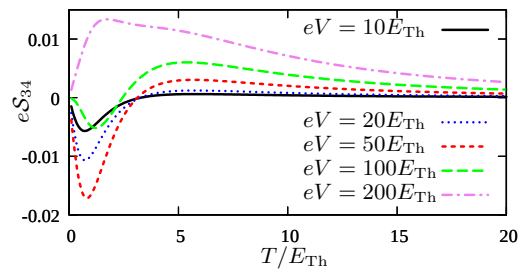}
\caption{The thermopower $\mathcal{S}_{34}$ for $\chi = 0$ as a function of temperature at different voltage bias values. Here we set $L_{S_1} = L_{S_2} = L_{N_1} = L_{N_2}=  1/3 L$, $L_{c,1} = 0.1 L$, $L_{N_3} = L_{N_4} = 1/2 L$ and $\Delta = 10^3E_{\mathrm{Th}}$.}
\label{fig: TempDep_phi_0}
\end{figure}

Making use of our general quasiclassical formalism one can evaluate the thermopower $\mathcal{S}_{34} (V, 0)$
numerically as a function of temperature. The results of this calculation are displayed in Fig.~\ref{fig: TempDep_phi_0}. We observe a rather non-trivial non-monotonous behavior of the thermopower as a function of $T$ which -- depending on the bias voltage -- can even change its sign. At high temperatures strongly exceeding the Thouless energy $E_{\mathrm{Th}}$ the thermopower $\mathcal{S}_{34}$ demonstrates a slow (power-law) decay with increasing $T$ which is reminiscent of that for the Aharonov-Bohm current component~\cite{GWZ97}.

In order to reconstruct the phase dependence of the thermoelectric signal it is necessary to resolve the kinetic equations \eqref{htusadel} and \eqref{hlusadel}. Then inside the normal wires $L_{N,3}$ and $L_{N,4}$ one finds
\begin{gather}
\begin{pmatrix}
I^T_{N_{3,4}} \\ I^L_{N_{3,4}} 
\end{pmatrix}
=
\hat{G}_{N_{3,4}}
\begin{pmatrix}
h^T_{N_{3,4}} - h^T_c \\ h^L_{N_{3,4}} - h^L_c
\end{pmatrix}.
\label{eqn: G_matrix}
\end{gather}
Here $h^{T/L}_c$ are the distribution function components evaluated in the central crossing point and $\hat{G}_{N_{3}}=\hat{G}_{N_{4}}  = \hat{G}_{N}$ is the spectral conductance matrix defined by Eq. \eqref{GN} with the integration performed either over the wire $L_{N_3}$ or $L_{N_4}$. In the absence of superconductivity this matrix reduces to the diagonal one proportional to the conductance of the corresponding normal wire. 

Bearing in mind that no current can flow into and out of electrically isolated terminals N$_{3}$ and N$_{4}$ and making use of the $T$-component of Eq.~\eqref{eqn: G_matrix}, we obtain
\begin{equation}
\int d \varepsilon
\Bigl\{
G^T_N [h^T_{N_3} -h^T_{N_4}]
+
G_N^{\mathcal{Y}} [h^L_{N_3} -h^L_{N_4}]
\Bigr\}
=0.
\end{equation}
This equation determines the relationship between the terminal temperatures $T_3,\ T_4$ and the induced thermoelectric voltages $V_3,\ V_4$. Assuming $\delta T = T_3 - T_4$ to be small enough and evaluating the  voltage difference $V_3-V_4$ up to linear in $\delta T$ terms, from Eq. \eqref{defS} we get
\begin{equation}
e\mathcal{S}_{34} \approx
\dfrac{1}{4T_N^2}
\dfrac{L_{N}}{\mathcal{A}\sigma_N}
\int
d \varepsilon
\dfrac{
(G^T_N + G^{\mathcal{Y}}_N) (\varepsilon + e V_N)
}{\cosh^2 [(\varepsilon + e V_N)/(2T_N)]}.
\label{s34}
\end{equation}
Here $V_N$ is the induced voltage in both terminals N$_3$ and N$_4$ for $T_3 = T_4 = T_N$, i.e. in the absence of the temperature gradient. It is also essential to keep in mind that the spectral conductance $G^T_N (\varepsilon, \chi)$ ($G^{\mathcal{Y}}_N (\varepsilon, \chi)$) in Eq. \eqref{s34} is an even (odd) function of both $\varepsilon$ and $\chi$.

The first conclusion to be drawn from the above result is that the thermoelectric effect vanishes in symmetric structures with $L_{S_1} = L_{S_2}$, $L_{N_1} = L_{N_2}$ and with the terminals N$_3$ and N$_4$ attached to the wire central point. In this case one has $V_N = 0$ and, hence, the even term in Eq.~\eqref{s34} vanishes after the energy integration. In addition, the spectral function $\mathcal{Y}$ equals to zero in the wires $L_{N_{3,4}}$, thus providing $G^\mathcal{Y} = 0$.
\begin{figure}
\centering
\includegraphics[width=80mm]{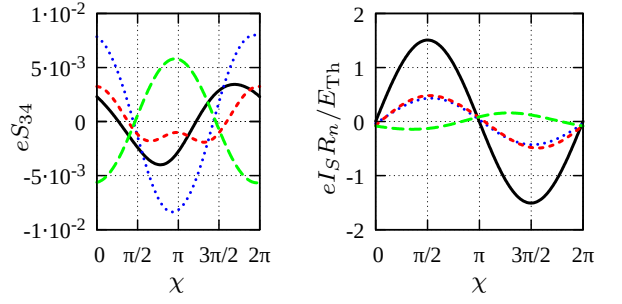}
\caption{The oscillating parts of the thermopower (left panel) and the current $I_S$ (right panel) as functions of $\chi$ at bias voltages $V$ equal to $10E_{\mathrm{Th}}/e$ (solid black lines), $50E_{\mathrm{Th}}/e$ (dotted blue lines), $100E_{\mathrm{Th}}/e$ (short dash red lines), $200E_{\mathrm{Th}}/e$ (long dash green lines).}
\label{fig: MainRes}
\end{figure}

In the general case the result \eqref{s34} can be expressed in the form \cite{DKZ18,dolgirev2018topology}
\begin{equation}
\mathcal{S}_{34} (V, \chi) = \mathcal{S}_0(V) + \mathcal{S}_{\mathrm{odd}}(V,\chi) + \mathcal{S}_{\mathrm{even}}(V,\chi),
\label{S34}
\end{equation}
where $\mathcal{S}_0(V) = \langle{\mathcal{S}}_{34}\rangle_\chi$ and the last two terms represent respectively odd and even in $\chi$ oscillating $2\pi$-periodic functions  with zero average over $\chi$. Hence, we may conclude that the thermopower $\mathcal{S}_{34} (V, \chi)$ \eqref{S34} is, in general, neither odd nor even function of $\chi$ taking a nonzero value at $\chi =0$.

At low enough voltages, such that $eV_N \lesssim E_{\mathrm{Th}}$ (which, however, does not necessarily imply $eV \lesssim E_{\mathrm{Th}}$), Eq.~\eqref{s34} implies that the odd in $\chi$ contribution dominates the thermopower.  At larger voltages $e V_N \gtrsim E_{\mathrm{Th}}$ the even harmonics starts to dominate, cf. Fig.~\ref{fig: MainRes}. The latter observation is specific to our geometry, where at such voltages the odd harmonics becomes suppressed by the factor $\sim (L_{c,1} - L_{c,2})/L$.

In Fig.~\ref{fig: MainRes} (left panel) we display the phase dependence of the thermopower at $T = E_{\mathrm{Th}}$ and for several voltage values $V$. We observe that, while at lower voltages the thermopower $\mathcal{S}_{34} (\chi)$ is (nearly) odd in $\chi$, it tends to an even function of $\chi$ quite rapidly with increasing $V$. 

Let us also compare this behavior of the thermopower with that of the phase-dependent current $I_S(\chi)$ between the two superconducting terminals. The corresponding results \cite{dolgirev2018interplay} are shown in the right panel of Fig.~\ref{fig: MainRes}.

In contrast to the thermopower $\mathcal{S}_{34} (\chi)$, the odd in $\chi$ harmonics dominates the function $I_S(\chi)$ at all given values of the bias voltage $V$. This observation allows to conclude that the thermopower $\mathcal{S}_{34} (\chi)$ may in general have a different origin from that of the current $I_S(\chi)$. This is because the current $I_S(\chi)$ is defined through the spectral supercurrent $j_E$, while the odd harmonics of $\mathcal{S}_{34} (\chi)$ originates from the spectral function $\mathcal{Y}$. We further notice that the temperature dependence of the odd current harmonics differs substantially from that of the thermopower: At sufficiently high temperatures $T\simeq 15E_{\mathrm{Th}}$ the former is essentially suppressed while the latter is not.


It is also worth pointing out that in the case of Andreev interferometers with a somewhat different topology from that analyzed here it was established ~\cite{DKZ18} that the thermopower-phase relation closely follows the dependence $I_S(\chi)$ at all values of $V$. This result seems to be in line with the work ~\cite{VH} where it was suggested that the thermopower $\mathcal{S}$ is simply proportional to ${\cal S} \propto d I_S/dT$. If so, one could assume that both the supercurrent and the thermopower have the same origin both being  odd functions of $\chi$. However, numerical calculations \cite{dolgirev2018topology} clearly {\it do not} support the relation $\mathcal{S}_{34} \propto d I_S/dT$ for Andreev interferometers under consideration.

Thus, we may conclude that the system topology -- along with such parameters as $V$, $T$ and $E_{\mathrm{Th}}$ -- plays a crucial role for the thermopower-phase relation. This conclusion qualitatively agrees with experimental observations \cite{Venkat1}. Depending on the system topology the behavior of the thermopower can be diverse: In symmetric setups it may become small or even vanish, while in non-symmetric ones the origin of the odd-part of the thermopower-phase relation can be determined either by the Josephson-like effect or by the spectral function $\mathcal{Y}$ which accounts for electron-hole asymmetry in our system.

\subsection{Long-range Josephson effect and $\pi$-junction states}
\label{subsec:longrange}
In Sec. \ref{subsec:ABJ} we demonstrated that the current flowing between superconducting terminals in the setup of Fig. \ref{snns3-fig} can be efficiently controlled biasing normal terminals by an external voltage $V=V_1-V_2$. Here we consider a different physical situation: The Andreev interferometer in Fig. \ref{snns3-fig} is exposed to a thermal gradient, i.e. normal terminals are kept at different temperatures $T_1$ and $T_2$
whereas no external voltage is applied $V=0$.

While this problem can be resolved in a general case \cite{KZ2021PRB}, the main physical effects are conveniently illustrated already
for a simpler geometry of Andreev interferometers with $L_p \to 0$ known as $X$-junctions \cite{KDZ20,KZ21EPJST}. 

To begin with, we consider an $X$-junction configuration with $L_{S_{1}}= L_{S_{2}}\equiv L/2$ and ${\mathcal A}_{S_{1}} ={\mathcal A}_{S_{2}}\equiv {\mathcal A}_{S}$.  In this particular case no electron-hole asymmetry is generated  \cite{KDZ20} and, hence, no thermoelectric effect occurs, i.e. $V_{1,2}=0$. Furthermore, the distribution function $h^T$ equals to zero, while the function $h^L$ inside the wires is defined by 
\begin{equation}
h^L = r^L_{N_2}h^L_{N_1}+r^L_{N_1}h^L_{N_2},
\label{hc}
\end{equation}
where $r^L_{N_i}=R^L_{N_i}/(R^L_{N_1} + R^L_{N_2})$ and 
\begin{equation}
R^L_{N_i} = 
\dfrac{1}{\mathcal{A}_{N_i}\sigma} \int\limits_{L_{N_i}} \dfrac{dx}{D^L}, \quad i=1,2.
\end{equation}

Let us identically rewrite Eq. \eqref{hc} in the form 
\begin{equation}
h^L = r_{N_2}h^L_{N_1} + r_{N_1}h^L_{N_2}+
W(\varepsilon) (h^L_{N_1} - h^L_{N_2}),
\label{hcW}
\end{equation} 
where we introduced the function $W(\varepsilon)=r^L_{N_2} r_{N_1} - r^L_{N_1} r_{N_2}$ which vanishes identically in structures with $L_{N_1} = L_{N_2} $ and remains nonzero otherwise decaying exponentially provided $|\varepsilon |$ exceeds the Thouless energy of our device $E_{\mathrm{Th}}$.

Making use of Eq. \eqref{hcW}, we immediately reconstruct the expression for the supercurrent $I_S$ between the superconducting terminals S$_1$ and S$_2$. We find \cite{KDZ20}
\begin{equation}
I_S = r_{N_2} I_J(T_1,\chi) + r_{N_1} I_J(T_2,\chi)+I_S^{\mathrm{ne}}(T_1,T_2,\chi),
\label{IS}
\end{equation}
where 
\begin{equation}
I_J(T,\chi)= - \dfrac{\sigma \mathcal{A}_S}{2e}\int j_{\varepsilon} \tanh \dfrac{\varepsilon}{2T}
d \varepsilon
\label{IJ}
\end{equation}
defines the equilibrium Josephson current and 
\begin{equation}
I_S^{\mathrm{ne}}=\dfrac{\sigma\mathcal{A}_S }{2e}
\int
j_{\varepsilon}
W ( \varepsilon)
\left(
\tanh \dfrac{\varepsilon}{2T_2} - \tanh \dfrac{\varepsilon}{2T_1}
\right)
d \varepsilon .
\label{ISne}
\end{equation}

Equations \eqref{IS}-\eqref{ISne} demonstrate that provided our Andreev interferometer is biased by a temperature gradient the supercurrent $I_S$  consists of two different contributions. The first one is a weighted sum of equilibrium Josephson currents $I_J$ \eqref{IJ} evaluated at temperatures $T_1$ and $T_2$ and the second one $I_S^{\mathrm{ne}}$ \eqref{ISne} accounts for non-equilibrium effects. 

Provided at least one of the two temperatures remains below the Thouless energy $E_{\mathrm{Th}}$, the current $I_S$ \eqref{IS} is dominated by the first (quasi-equilibrium) contribution, while the non-equilibrium one \eqref{ISne} can be ignored. In the opposite limit $T_{1,2} \gg E_{\mathrm{Th}}$, the equilibrium contribution to $I_S$ is exponentially suppressed and reads \cite{KDZ20}:
\begin{multline}
I_J= \dfrac{16 \varkappa}{ 3 + 2\sqrt{2}}\dfrac{E_{\mathrm{Th}}}{eR_n}\left(\dfrac{2 \pi T}{E_{\mathrm{Th}}}\right)^{3/2}
e^{-\sqrt{2 \pi T /E_{\mathrm{Th}}}} \sin\chi 
\label{IJeq}
\end{multline}
where $R_n=L/(\mathcal{A}_S\sigma )$ and the parameter $\varkappa = 4 \sqrt{\mathcal{A}_{S_1}\mathcal{A}_{S_2}}/(\mathcal{A}_{S_1} + \mathcal{A}_{S_2} + \mathcal{A}_{N_1} + \mathcal{A}_{N_2})$ is taken here at ${\mathcal A}_{S_{1}} ={\mathcal A}_{S_{2}}$. Hence, the supercurrent can now be dominated by the non-equilibrium term  
\begin{multline}
I_S^{\mathrm{ne}} \simeq
0.21\varkappa^3 r_{N_1} r_{N_2}\dfrac{ E^2_{\mathrm{Th}}}{eR_n}\left(\dfrac{1}{T_1} - \dfrac{1}{T_2} \right)\\\times\left(\dfrac{L}{L_{N_2}}-\dfrac{L}{L_{N_1}}\right) 
\sin \chi \cos^2(\chi/2).
\label{Wapprox}
\end{multline}

\begin{figure}
\centering
\includegraphics[width=80mm]{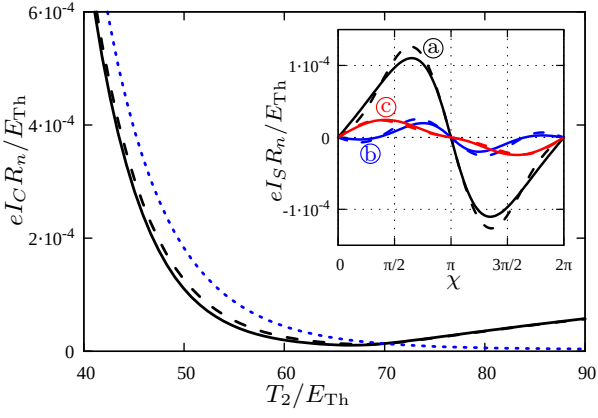}
\caption{Josephson critical current $I_C\equiv {\mathrm{max}}|I_S|$ as a function of $T_2$. Inset: CPR evaluated at $T_2= 50 E_{\mathrm{Th}}$ (a),  $60 E_{\mathrm{Th}}$ (b) and $75 E_{\mathrm{Th}}$ (c). Solid lines correspond to the exact numerical solution, dashed lines indicate the result \eqref{IS} combined with \eqref{IJeq} and \eqref{Wapprox}, dotted line is the quasi-equilibrium contribution $r_{N_2} I_J(T_1,\pi/2) + r_{N_1} I_J(T_2,\pi/2)$ to $I_S$. The parameters are: $T_1 = 70 E_{\mathrm{Th}}$, $L_{S_{1,2}}=L/2$, $L_{N_1}=3 L$, $L_{N_2}=L$ and $\mathcal{A}_{S_1} = \mathcal{A}_{S_2} = \mathcal{A}_{N_1} = \mathcal{A}_{N_2}$. }
\label{Ic-T-chi-3-1-fig}
\end{figure}

\begin{figure}
\centering
\includegraphics[width=80mm]{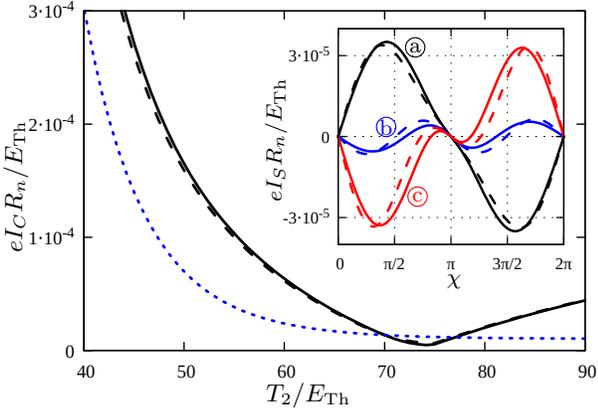}
\caption{The same as in Fig. \ref{Ic-T-chi-3-1-fig}. The parameters are the same except $L_{N_1}=L$, $L_{N_2}=3 L$. Temperature values in the inset are $T_2= 65 E_{\mathrm{Th}}$ (a),  $75 E_{\mathrm{Th}}$ (b) and $85 E_{\mathrm{Th}}$ (c).}
\label{Ic-T-chi-1-3-fig}
\end{figure}

We observe that at temperatures strongly exceeding the Thouless energy, the supercurrent $I_S\simeq I_S^{\mathrm{ne}}$ decays as a power-law with increasing min$(T_1,T_2)$, unlike the equilibrium Josephson current in long SNS junctions which is known to decay exponentially. This behavior is due to driving the electron distribution function  $h^L$ out of equilibrium by exposing the system to a temperature gradient. In Figs. \ref{Ic-T-chi-3-1-fig} and \ref{Ic-T-chi-1-3-fig} we display the critical Josephson current $I_C$ as a function of $T_2$ for fixed $T_1$. At high temperatures, $T_2 \gg E_{\mathrm{Th}}$ the current $I_C$ strongly exceeds both equilibrium values $I_J(T_1)$ and $I_J(T_2)$ and even starts to grow for $T_2\gtrsim T_1$. This behavior implies {\it strong supercurrent stimulation by a temperature gradient}.

Another remarkable feature of our result \eqref{Wapprox} is the non-sinusoidal CPR that persists even at temperatures strongly exceeding
$E_{\mathrm{Th}}$. Note that the dependence of the equilibrium Josephson current on the phase $\chi$ in SNS junctions remains 
non-sinusoidal only at $T \lesssim E_{\mathrm{Th}}$ and reduces to $I_J \propto \sin \chi$ at higher temperatures.

In addition, we observe that the sign of the supercurrent in Eq. \eqref{Wapprox} is controlled by those of both length and temperature differences, $L_{N_1} - L_{N_2}$ and $T_1 - T_2$. For instance, by choosing $L_{N_1} < L_{N_2}$ and $T_1 <T_2$ we arrive at a pronounced $\pi$-junction-like behavior, see also Fig. \ref{Ic-T-chi-1-3-fig}.

\begin{figure}
\centering
\includegraphics[width=70mm]{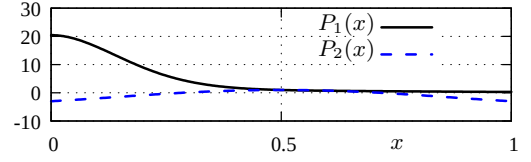}
\caption{Universal functions $P_{1}(x)$and $P_{2}(x)$. 
We observe that $P_1(1/2) = P_2(1/2) = 1$.}
\label{P-fig}
\end{figure}

For fully asymmetric X-junction supercurrent $I_S$ can be also represented in the form \eqref{IS} with lengthy expression for non-equilibrium term \cite{KZ21EPJST}. Our numerical calculation demonstrates that with a good accuracy non-equilibrium contribution can be approximately represented in the form \eqref{ISne} with functions $j_{\varepsilon}$ and $W(\varepsilon)$ evaluated for particular asymmetric X-junction.

Evaluating these functions in the high energy limit and extrapolating them to the whole energy interval one can analytically evaluate non-equilibrium current at sufficiently high temperatures $T_{1,2} \gg E_{\mathrm{Th}}$ which reads
\begin{multline}
I_S^{\mathrm{ne}} = 
\dfrac{4\varkappa^3}{(3+2\sqrt{2})^2} \dfrac{1101}{1250}
r_{N_1} r_{N_2}
\left(\dfrac{1}{T_1} - \dfrac{1}{T_2}\right)
\dfrac{ E_{\mathrm{Th}}}{eR_n^a}
\times\\
\left(
\dfrac{L}{L_{N_2}} - \dfrac{L}{L_{N_1}}
\right)
\biggl\{
\dfrac{\mathcal{A}_{S_1}}{2 \mathcal{A}_{S_2}} P_1(L_{S_1}/L)
+\\
\dfrac{\mathcal{A}_{S_2}}{2 \mathcal{A}_{S_1}}  P_1(L_{S_2}/L)
+
P_2(L_{S_1}/L)
\cos \chi
\biggr\}
\sin \chi ,
\label{Ktheor}
\end{multline}
where the universal functions $P_1$ and $P_2$ are displayed in Fig. \ref{P-fig}. This result demonstrates that in a wide temperature interval $T_{1,2}>E_{\mathrm{Th}}$ the non-equilibrium contribution to the Josephson current  is described by a universal power law dependence $\propto 1/T_1 - 1/T_2$, whereas the phase dependence of $I_S^{\mathrm{ne}}$ depends on the junction geometry only. 

For partially symmetric junction with $L_{S_1}=L_{S_2}$ and $\mathcal{A}_{S_1} = \mathcal{A}_{S_2}$ we have $P_1=P_2=1$ and the phase dependence of $I_S^{\mathrm{ne}}$ reduces to the form \eqref{Wapprox}. For strongly asymmetric junctions the function $P_1$ increases by about an order of magnitude, whereas $P_2$ varies slightly (see Fig. \ref{P-fig}). In this case the current-phase relation approaches a sin-like form. Numerical calculation shows that our analytic formula \eqref{Ktheor} is in a good agreement with numerically exact results for $I_S^{\mathrm{ne}}$  as long as the lengths $L_{S_{1,2}}$ remain not very small $\min(L_{S_1}, L_{S_2}) \gtrsim 0.2 L$.

\begin{figure}
\centering
\includegraphics[width=70mm]{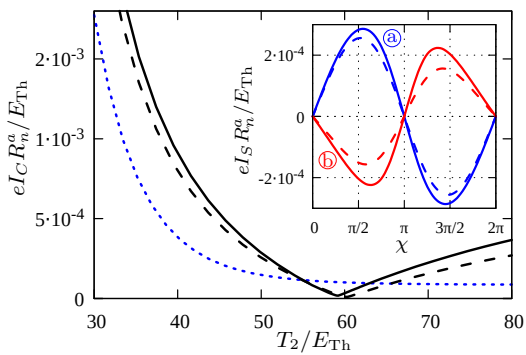}
\caption{Josephson critical current $I_C\equiv \max |I_S|$ as a function of $T_2$. Inset: The phase dependencies of the Josephson current $I_S(\chi)$ for $T_2 = 50 E_{\mathrm{Th}}$ (``a'' curves) and $T_2 = 70 E_{\mathrm{Th}}$ (``b'' curves).
Solid lines correspond to our numerically exact solution, dashed lines indicate the result \eqref{IS} combined \eqref{Ktheor}, dotted line represent the quasi-equilibrium contribution $r_{N_2} I_J(T_1,\pi/2) + r_{N_1} I_J(T_2,\pi/2)$ to $I_S$. 
The parameters are: $T_1=55 E_{\mathrm{Th}}$, $L_{N_1}=L$, $L_{N_2}=3 L$, $L_{S_1}=0.2L$ and $\mathcal{A}_{S_1} = \mathcal{A}_{S_2} = \mathcal{A}_{N_1} = \mathcal{A}_{N_2}$.}
\label{ISc-LS1-2-LS2-8-LN1-1-LN2-3-fig}
\end{figure}

\begin{figure}
\centering
\includegraphics[width=80mm]{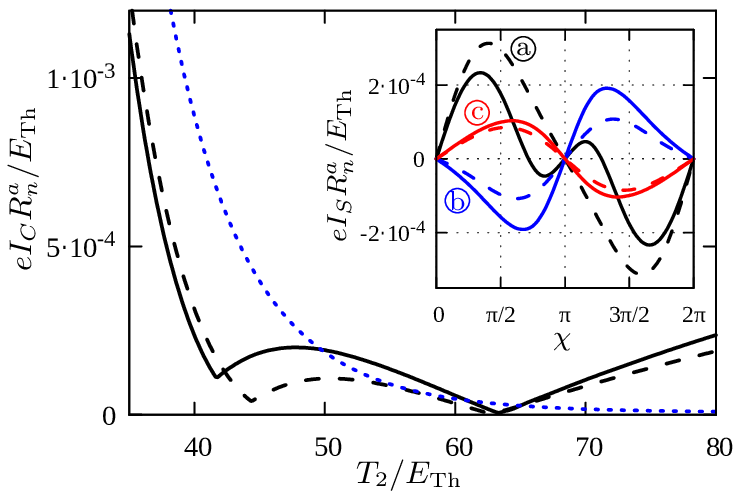}
\caption{The same as in Fig. \ref{ISc-LS1-2-LS2-8-LN1-1-LN2-3-fig}. The parameters are the same except $L_{N_1}=3 L$, $L_{N_2}=L$ and $T_1=65 E_{\mathrm{Th}}$.
Temperature values in the inset are $T_2= 40 E_{\mathrm{Th}}$ (a),  $50 E_{\mathrm{Th}}$ (b) and $70 E_{\mathrm{Th}}$ (c).}
\label{ISc-LS1-2-LS2-8-LN1-3-LN2-1-fig}
\end{figure}

It is easy to verify that the term $I_S^{\mathrm{ne}}$ dominates the supercurrent $I_J$ \eqref{IJeq} already at  $\min T_{1,2} \gtrsim 30E_{\mathrm{Th}}$, i.e. at considerably lower temperatures than in symmetric junctions considered above, where the analogous condition reads $T_{1,2} \gtrsim 70 E_{\mathrm{Th}}$. In Figs. \ref{ISc-LS1-2-LS2-8-LN1-1-LN2-3-fig} and \ref{ISc-LS1-2-LS2-8-LN1-3-LN2-1-fig} we also observe that for asymmetric junctions with  $L_{S_1} \ll L_{S_2}$ and $\mathcal{A}_{S_1} = \mathcal{A}_{S_2} = \mathcal{A}_{N_{1}} = \mathcal{A}_{N_{2}}$ non-equilibrium effects are visible only at $T_{1,2}/E_{\mathrm{Th}}\gtrsim 40\div 50 $. Hence, we conclude that strongly asymmetric $X$-junctions with $L_{S_1} \ll L_{S_2}$ and $\mathcal{A}_{S_1} \gg \mathcal{A}_{S_2} + \mathcal{A}_{N_1}+ \mathcal{A}_{N_2}$ appear to be most suitable candidates for observing the non-equilibrium Josephson current (\ref{Ktheor}).

Finally, we remark that for $L_{N_1} > L_{N_2}$ our $X$-junction may exhibit two transitions between $0$- and $\pi$-junction states. In  this case the system is in the 0-junction state as long as $T_2$ remains low enough to keep the quasi-equilibrium term larger than  $I_S^{\mathrm{ne}}$. However, since with increasing $T_2$ (albeit for $T_2< T_1$)  the contribution $\propto I_J$ decays faster than the non-equilibrium one (now having a negative sign), the $X$-junction eventually switches to the $\pi$-junction state. Further increasing $T_2$ one reachs the point  $T_2=T_1$ where  $I_S^{\mathrm{ne}}$ changes its sign, thus signaling the transition back to the $0$-junction state at $T_2$ slightly below $T_1$. This behavior is illustrated in Fig. \ref{ISc-LS1-2-LS2-8-LN1-3-LN2-1-fig}.

\section{Concluding remarks}
\label{sec:remarks}

For decades thermoelectricity in superconductors was and remains one of the most intriguing topics of modern condensed matter physics \cite{NL}. In this review we outlined some recent developments in the field focusing our attention on multi-terminal superconducting-normal hybrid structures. We demonstrated that electron-hole symmetry breaking in such systems may cause enormous enhancement of thermoelectric effects and elucidated several physical mechanisms for such symmetry breaking. Phase-coherent nature of thermoelectric effects manifests itself in a periodic dependence of the thermopower on the applied magnetic flux indicating their close relation to Josephson and Aharonov-Bohm effects. We also demonstrated that dc Josephson current in multi-terminal hybrid structures can be efficiently tuned by exposing the system to a temperature gradient leading to a nontrivial current-phase relation and to a possibility for $\pi$-junction states. 

Finally, it is worth pointing out that thermoelectric effects in superconductors give rise to a variety of applications ranging from thermometry and refrigeration \cite{Pekola} to phase-coherent caloritronics \cite{Giazotto} that paves the way to an emerging field of
thermal logic \cite{Li} operating with information in the form of energy. We hope that theoretical results and predictions discussed in this work
not only shed light on some previously unresolved issues but also could help to put forward these and other applications of thermoelectric effects.

\end{document}